# Direct measurement of electrocaloric effect based on multi-harmonic lock-in thermography

Ryo Iguchi,[1,a)] Daisuke Fukuda,[2] Jun Kano,[2] Takashi Teranishi,[2,3,a)] and Ken-ichi Uchida,[1,4]

[1]*National Institute for Materials Science, Tsukuba 305-0047, Japan*

[2]*Graduate School of Natural Science and Technology, Okayama University, Okayama 700-8530, Japan*

[3]*Laboratory for Materials and Structures, Tokyo Institute of Technology, Yokohama 226−8503, Japan*

[4]*Institute for Materials Research, Tohoku University, Sendai 980-8577, Japan*

[b)] Author to whom correspondence should be addressed: IGUCHI.Ryo@nims.go.jp and terani-t@cc.okayama-u.ac.jp

**Abstract:** In this study, we report on a direct measurement method for the electrocaloric effect, the heating/cooling upon application/removal of an electric field in dielectric materials, based on a lock-in thermography technique. By use of sinusoidal excitation and multi-harmonic detection, the actual temperature change can be measured by a single measurement in the frequency domain even when the electrocaloric effect shows nonlinear response to the excitation field. We have demonstrated the method by measuring the temperature dependence of the electric-field-induced temperature change for two Sr-doped BaTiO$_3$ systems with different ferroelectric-paraelectric phase transition temperatures, where the procedure for extracting the pure electrocaloric contribution free from heat losses and Joule heating due to leakage currents is introduced. This method can be used irrespective of the type of dielectric materials and enables simultaneous estimation of the polarization change and power dissipation during the application of the electric field, being a convenient imaging measurement method for the electrocaloric effect.

The electrocaloric effect (ECE) refers to the temperature change as a result of the application and removal of an electric field in dielectrics. Since the observation of giant ECE-induced temperature change,[1–4] it gains considerable attention for energy-efficient and environment-friendly temperature modulators without greenhouse gases. To realize thermal management applications based on ECE, intense efforts have been devoted for both materials research and engineering.[5–9]

To investigate the performance of ECE, it is necessary to know how large temperature change is generated for the simplest cycle of applying and removing the electric field *E*. To do this, there are two known approaches: indirect and direct methods.[10–12] The indirect method is based on the Maxwell's relation; the temperature $T$ dependence of the electric polarization $P$ at various values of $E$ is converted into the adiabatic temperature change $\Delta T_{\text{ECE}}^{\text{ad}}$ induced by ECE through $\Delta T_{\text{ECE}}^{\text{ad}} = \int_0^{E_{\max}} C_v^{-1} T (\partial P/\partial T)_E \, dE$ with $C_v$ and $E_{\max}$ respectively being the volumetric specific heat and maximum electric field applied to a dielectric material. This method, however, cannot be applied to the systems that show spatially nonuniform polarization texture:[7,13] e.g., domain structures in ferroelectrics and polar nano regions due to short-range ordering in relaxer-type ferroelectrics. Another problem in the indirect method is that the $E$ dependence of $C_v$ is often neglected despite its considerable contribution.[10,11,14] In contrast, the direct method is free from these problems because $\Delta T_{\text{ECE}}^{\text{ad}}$ is directly measured by applying and removing $E$ in a quasi-adiabatic condition using thermometers or calorimeters. The only drawback in the direct method is the difficulty for suppressing heat losses between samples and sensors and from samples to environment while applying $E$.[10] From these aspects, the non-contact direct measurements using thermography has been used recently.[8–10,15,16]

In this study, we propose and demonstrate the direct ECE-measurement method based on the lock-in thermography (LIT) technique.[17–20] In LIT, for enhancing the signal-to-noise ratio, AC excitation is applied to a sample, and the coupled AC temperature change response is extracted through Fourier analysis. LIT should be beneficial for ECE studies as it has been used for measuring the temperature change induced by the magnetocaloric and elastocaloric effects.[21–23] However, the conventional LIT method is applicable only to the linear response regime, while most of the caloric effects are intrinsically nonlinear.[24–26] Thus, in the previous study, the integration of the temperature change measured with small amplitude excitation at various bias fields is required for quantitative measurements.[21] Such a situation is inconvenient if samples are ferroic and show hysteresis during the excitation. To overcome this limitation, we have extended the LIT technique with introducing multi-harmonic detection with sinusoidal excitation (Fig. 1) and demonstrated one-shot quantitative measurement of $\Delta T_{\text{ECE}}^{\text{ad}}$. Here, we measure the temperature change induced when $E$ is varied from 0 to $E_{\max}$: $\Delta T_{\max}$, which is connected to $\Delta T_{\text{ECE}}^{\text{ad}}$. The sinusoidal excitation also allows tracking of the polarization change and associated electrical consumption during the ECE excitation. In the following, we first explain the experimental procedure and then demonstrate it by measuring the $T$ dependence of ECE in Sr-substituted BaTiO$_3$. We used two samples with different compositions, Ba$_x$Sr$_{1-x}$TiO$_3$ (BST) with $x$=0.6 and $x$=0.8, respectively showing paraelectricity and ferroelectric-paraelectric phase transition in the measurement temperature range.

In the LIT measurements of the ECE-induced temperature change, the sinusoidally oscillating electric field $E(t) = E_{\max}[\sin(2\pi f t)/2 + 1/2]$ is applied to samples instead of the step-function like excitation used in the time-domain measurements.[8,15,27] Here, $f$ and $t$ denotes the excitation frequency and time, respectively. The ECE-induced temperature change should take the maximum at $E(t) = E_{\max}$ when the relaxation is fast so that the dynamic $P$ value follows the static $P$-$E$ curve. The temperature change between the zero field and $E_{\max}$ is given by

$$\Delta T_{\max} = 2 \sum_{n \in \text{odd}} (-1)^{\frac{n-1}{2}} A_{nf} \cos(\phi_{nf}), \qquad (1)$$

where $A_{nf}$ and $\varphi_{nf}$ denotes the amplitude and phase at $n$-th higher harmonic frequencies ($nf$), respectively. Thus, to estimate $\Delta T_{\max}$, the measurements not only at the 1st harmonic but also at odd higher harmonics are necessary [Fig. 1(b)]. Note that when the waveform of $E(t)$ is cosine, $(-1)^{(n-1)/2}$ is not necessary. $\Delta T_{\max}$ becomes equivalent to $\Delta T_{\text{ECE}}^{\text{ad}}$ when the contributions due to heat losses[28] and the Joule heating, induced by leakage charge currents in dielectric materials,[27] are suppressed. Such a condition can be realized at sufficiently high excitation frequencies and can be confirmed by measuring the $f$ dependence of the signal as follows. The ECE heating/cooling power increases with increasing $f$, and the resultant temperature change is constant over $f$. The heat losses reduce the temperature changes due to ECE and Joule heating by breaking the system adiabaticity but are remarkable only at low $f$. In contrast, the Joule heating power [$\propto E(t)^2$] is constant over $f$, and the resultant temperature change decreases with increasing $f$ (see Supplementary Material for the mathematical model). Therefore, the measurements at sufficiently high $f$ allow the estimation of the pure ECE-induced contribution. We note that, in actual measurements, the dielectric loss heating due to the imaginary part of the permittivity or hysteresis inevitably appears as in time-domain measurements and may contribute to $\Delta T_{\text{max}}$. However, its effect is expected to be minor; the dielectric loss heating contribution mainly appears at the even-harmonic frequencies because it occurs regardless of increasing or decreasing $E$, while only the difference depending on the $E$-sweep direction contributes to $\Delta T_{\text{max}}$.

The LIT measurements were performed with a microbolometer-based infrared camera with a detection wavelength range of 7.5 to 14.0 µm and a pixel number of 1024 × 96 at a framerate of 240 Hz, where one pixel corresponds to an area of 35×35 µm$^2$. The thermal images were continuously and periodically captured with its timing used as a clock for generating a reference sinusoidal wave with the frequency $f$ by a timing controller [Fig. 1(a)]. The captured thermal images were transferred to a computer for calculating the amplitude $A_{nf}$ and phase $\varphi_{nf}$ images at the harmonic frequency $nf$ based on Fourier analysis [Fig. 1(b)]. Here, the $A_{nf}$ images represent the distribution of the magnitude of the temperature change and the $\varphi_{nf}$ images represent its sign (heating or cooling) and delay due to thermal conduction. Note that the maximum valid lock-in frequency is the quarter of the camera's framerate (60 Hz in this study), and the maximum $nf$ value is limited accordingly. The obtained images were calibrated for compensating the material-dependent infrared emissivity and the time constant of the microbolometer sensors. All the measurements were performed at atmospheric pressure.

The BST polycrystalline samples were prepared by conventional solid-state reaction from powders of BaCO$_3$, SrCO$_3$, and TiO$_2$ with changing the stoichiometric ratio for varying composition following to the previous study[29]. For the ECE measurements, the obtained pellets were cut into a rectangular slab of 2.0×2.0×1.0 mm$^3$ using a diamond wire saw, followed by the mechanical polishing. On the 2.0×2.0 mm$^2$ surfaces, Au electrodes with a thickness of ~100 nm were sputter-deposited, and subsequently Au wires with a diameter of 0.05 mm were attached to the electrodes using a small amount of silver paste. Two $x$=0.6 (labeled as #1 and #2) and two $x$=0.8 (#3 and #4) samples were placed onto a Peltier-module-based temperature control stage, where the samples and stage are electrically and thermally insulated by inserting a 0.5-mm-thick sapphire plate and 0.15-mm-thick double-sided polyimide tape. For the calibration, dummy samples of which the half of the top surface was covered by black ink were located near the main samples [see samples at the left of #1 and #3 in Fig. 1(c)]. We confirmed that the difference between the estimated temperatures for the bare and black-ink-coated surfaces is less than 1%, indicating the sufficiently large infrared emissivity of the BST samples themselves. Each sample was connected to a high voltage amplifier with a Sawyer-Tower circuit[30] consisting of a parallel connection of a reference capacitor of 0.1 µF and a resistance of 10 MΩ for DC bias stabilization. Using the Sawyer-Tower circuit, we can monitor the $P$-$E$ curve of the BST samples during the LIT measurement. Figure 2(a) shows that the $x$=0.6 and $x$=0.8 samples respectively exhibit paraelectricity and ferroelectricity at room temperature, where the accumulated charge $\delta Q$ in the reference capacitor and the corresponding polarization change $\delta P$ were calculated after subtracting the contribution from the cable parasitic capacitance. The dielectric maximum temperature $T_m$ for the $x$=0.6 ($x$=0.8) samples was 270 K (350 K).[31]

Now we show the results of the LIT measurements. We first focus on the relative behavior of the first harmonic temperature change, i.e., $A_{1f}$ and $\varphi_{1f}$, and measured their $f$ dependence to confirm the heat losses and Joule heating contributions. Figure 2(b) shows the $A_{1f}$ and $\varphi_{1f}$ signals for the $x$=0.6 and $x$=0.8 samples at $E_{max}$=1.0 kV/mm at various temperatures, extracted from 25×55 pixels inside the sample surfaces [see white dashed rectangles in Fig. 1(c)]. We found that the extracted $A_{1f}$ values are constant for $f \geq 10$ Hz and $\varphi_{1f}$ values converge to ~ 0°, indicating that the sign of the observed temperature modulation coincides with that due to ECE (heating when applying $E$ and cooling when removing $E$) in both the samples and that $f \geq 10$ Hz is the sufficient condition to assure the system adiabaticity and separation from Joule heating. This behavior was obtained at various temperatures, and hereafter we focus the results at $f$=10 Hz. We also confirmed the nonlinear nature of ECE in our samples by measuring the $E_{max}$ dependence of $A_{1f}$ and $\varphi_{1f}$. As shown in Fig. 2(c), the observed $A_{1f}$ contribution increases with increasing $E_{max}$ almost quadratically, although it is not the whole of the ECE contribution due to the presence of the higher-harmonic contribution. The $\varphi_{1f}$ values are unchanged from ~0°, indicating the dominant contribution of ECE over the measurement $E$ range as the Joule heating contribution shows $\varphi_{1f}=90°$ for the adiabatic condition (see Supplementary Material).[32]

Next, we extract $\Delta T_{\max}$ through the multi-harmonic detection. As shown in Fig. 2(c), ECE nonlinearly depends on $E$ and may induce higher harmonic temperature changes, so that these contributions should be taken into consideration for deducing $\Delta T_{ECE}^{ad}$. Figure 3(a) shows the harmonic $nf$ and temperature $T$ dependences of the in-phase signal $A_{nf}\cos(\phi_{nf})$ and quadrature signal $A_{nf}\sin(\phi_{nf})$, where only the former contributes to $\Delta T_{\max}$. It is found that the magnitude of the signals with $nf>1$ is much smaller than that of the $1f$ signals in our samples. The highest contribution can be found for $A_{3f}\cos(\phi_{3f})$ at $T\sim T_m$ for the $x=0.8$ samples. Figure 3(b) shows $\Delta T_{\max}$ as a function of $T$, calculated based on Eq. (1). The magnitude of $\Delta T_{\max}$ monotonically decreases with increasing $T$ for $x=0.6$, while it shows a peak around $T_m$ (~ 350 K) for $x=0.8$. For each composition, the two samples (#1-2 for $x=0.6$ and #3-4 for $x=0.8$) show the same tendency, confirming the high reproducibility of our results.

The observed results are consistent with the expected behavior of ECE that the induced temperature modulation is maximized around $T_m$.[7,12] As $T_m$ for the $x=0.6$ samples (#1-2) is below our measurement range, $\Delta T_{\max}$ increases with decreasing $T$. Furthermore, since the samples exhibit a paraelectric behavior with cubic symmetry in the measurement $T$ range, $\Delta T_{\max}$ can be quantitatively estimated by the indirect measurement. The yellow solid line in Fig. 3(b) represents the $\Delta T_{ECE}^{ad}$ indirectly estimated by assuming $C_V=3.06\times10^6$ J/K/m$^3$ based on a typical value for BaTiO$_3$ at 300 K[33] and disregarding its $T$ and $E$ dependences for simplicity. The calculation well reproduces the value and tendency of the ECE signals observed directly. The observed behavior for the $x=0.8$ samples (#3-4) that $\Delta T_{\max}$ is maximized at $T_m$ is also as expected since they show the ferroelectric-paraelectric phase transition at $T_m$. Different from the $x=0.6$ samples, the quantitative estimation based on the indirect method is difficult because the cycle of $E=0$ to $E_{max}$ corresponds to the minor loop operation in the ferroelectric phase and thus includes domain dynamics, which prohibits the simple application of the Maxwell's relation. Indeed, as shown in Fig. 2(a), the polarization change $\delta P$ during the ECE measurements (shown as the solid line) is much smaller than that for the symmetric cycle between $-E_{max}$ and $+E_{max}$ (shown as the dashed line). Note that as the polarization depends nonlinearly on $E$, $\Delta T_{max}$ is more affected by the odd-number harmonic signals ($nf=3, 5, 7…$) [Fig. 3(a)]. Therefore, the multi-harmonic detection is especially important for investigating ECE around $T_c$.

The sinusoidal excitation allows the simultaneous characterization of the electrical properties of the samples during the LIT measurements. Figure 3(c) shows the DC and AC power consumptions, $P_{DC}$ and $P_{AC}$, for driving ECE, estimated using the incorporated Sawyer-Tower circuit. $P_{DC}$ is calculated from the DC voltage and current applied to the samples (measured by the resistor in the circuit) and $P_{AC}$ is calculated by averaging the AC current (measured by the reference capacitor) multiplied by the AC voltage over one period. Since $P_{AC}$ is proportional to the number of the cycles per second (i.e., $f$), $P_{AC}/f$ is plotted. $P_{DC}$ is found to monotonically increase as $T$ increases for all the samples, which shows that the leakage current increases with $T$. This is consistent with the increased Joule heating contribution at high $T$ in the LIT data [see the behaviors of the temperature change in the low $f$ region in Fig. 2(b)]. On the other hand, $P_{AC}$ drastically depends on the composition. The $P_{AC}$ value for the $x=0.6$ samples shows small increase with $T$, which is attributed to the electrical leakage. In contrast, the $P_{AC}$ value for the $x=0.8$ samples is much larger at room temperature and decreases with increasing $T$ after taking a peak around 350 K. This behavior is attributed to the phase transition from a ferroelectric to paraelectric state around $T_c$, where the hysteresis loss originated from domain fluctuation disappears. The effect of the electrical leakage in $P_{AC}$ for the $x=0.8$ samples can be found at high temperatures (e.g., $T>360$ K), where $P_{AC}$ starts to increase with increasing $T$ in a similar manner to $P_{DC}$. We note that the electrical leakage also degrades the estimation of $\delta P$ from $\delta Q$ [Fig. 2(a)] because not only the polarization changes but also the leakage current contributes to $\delta Q$, even though $\delta Q$ itself and power consumptions are correctly estimated.

Finally, we discuss the benefits provided by the LIT-based direct measurement method for ECE. Thermography enables non-contact direct temperature measurements, and its imaging capability allows one to measure the characteristics of ECE in multiple samples simultaneously, leading to reliable and quick materials screening.[21,22] The proposed method can skip integration of the data observed at various bias fields and can be used even in the ferroelectric phase, further improving the measurement throughput. Significantly, the imaging capability also identifies the non-uniformity of the temperature change due to ECE. In general, the LIT measurements at higher $f$ can reveal heat source distributions due to the decrease of the thermal diffusion length.[17,34] Figure 4(a) [4(b)] compares the $\Delta T_{max}$ images measured at low and high $f$ values at around 295 K (385 K). It can be found that the high frequency excitation results in the localized $\Delta T_{max}$ distribution, which illustrates the composition and/or defect distributions. The prominent localization was observed for the $x$=0.8 samples (#3-4) at room temperature, indicating that the ferroelectricity appears non-uniformly with creating domains that may be coupled to the sample non-uniformity. In fact, the non-uniformity of $\Delta T_{max}$ for the $x$=0.8 samples disappears in the paraelectric phase ($T$>360 K), highlighting the role of ferroelectric domain structures.

In summary, we demonstrated the LIT-based direct measurement method for ECE. Compared with the previous measurements of the caloric effects, the sinusoidal waveform of the applied electric field and multi-harmonic detection enable the one-shot measurement of the ECE-induced temperature change, which is in principle applicable to other caloric effects including multicaloric responses.[4,35] Furthermore, we show that the detection of the polarization change and estimation of the power dissipation can be readily achievable during the operation of ECE. We anticipate that rich information extracted by the proposed method accelerates searching of ECE materials and contributes to developing solid-state thermal management devices.

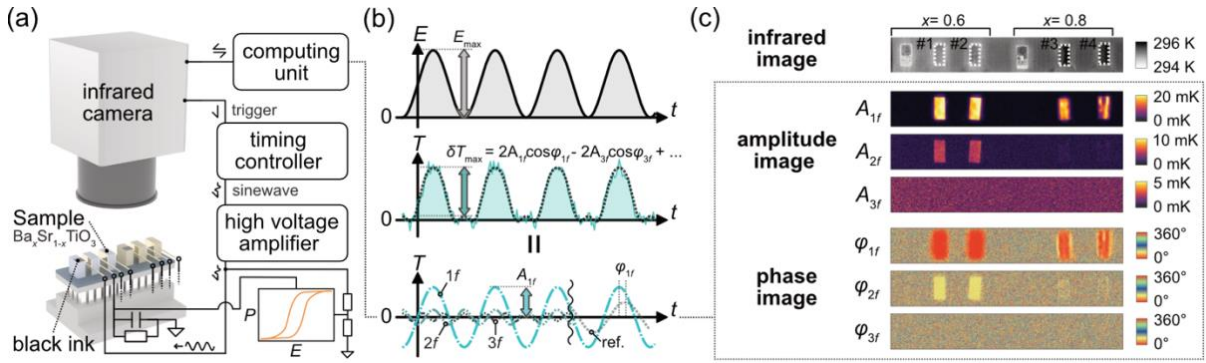

Fig. 1 (a) Experimental setup for the lock-in thermography measurement. Each sample is connected to a voltage source in parallel via a Sawyer-Tower circuit for measuring the polarization $P$ versus the excitation field $E$. To generate $E$, the reference sinusoidal wave is inputted to the high voltage amplifier. (b) Time $t$ dependence of the excitation field $E$ and temperature $T$. Decomposition of multi-harmonic signals for detecting the maximum temperature change is also shown. (c) Raw infrared images before the calibration (top panel) and lock-in amplitude $A_{nf}$ and phase $\varphi_{nf}$ images at different harmonic frequency $nf$ (bottom). The images are cropped to 700×96 pixels. The dotted white rectangle indicates the area (25×55 pixels) used for calculating the averaging values.

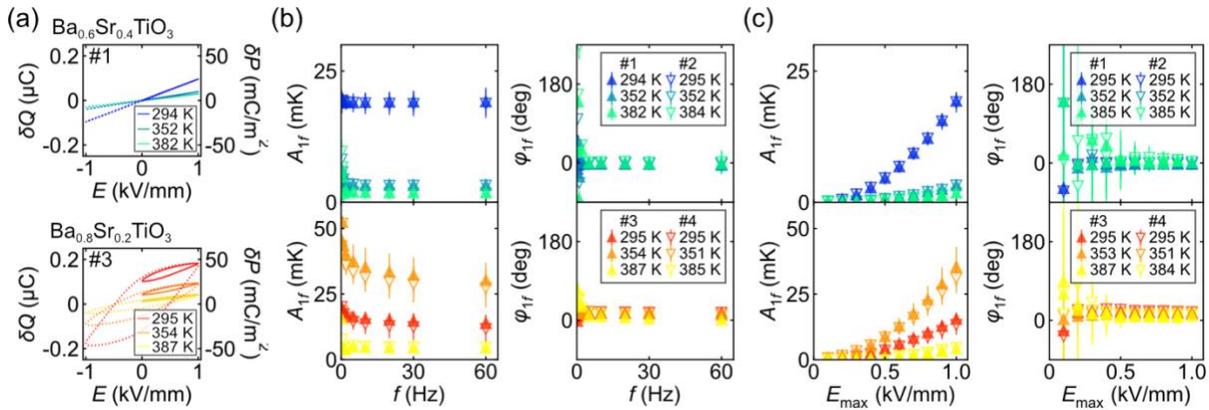

Fig. 2 (a) Accumulated charge $\delta Q$ in the reference capacitor and corresponding polarization change $\delta P$ as a function of $E$. The solid (dotted) lines represent the results for the 0 (-$E_{max}$) to +$E_{max}$ (+$E_{max}$) cycle, where the offset is added for the maximum (average) value to coincide with the maximum value for the -$E_{max}$ to +$E_{max}$ cycle (to be zero). (b) Frequency $f$ dependence of the 1st harmonic amplitude $A_{1f}$ and phase $\varphi_{1f}$ at various temperatures. (c) $E_{max}$ dependence of $A_{1f}$ and $\varphi_{1f}$ at various temperatures. The $T$, $A_{1f}$, and $\varphi_{1f}$ values were obtained by averaging the 25×55 values inside the samples [see the rectangular box in Fig. 1(c)].

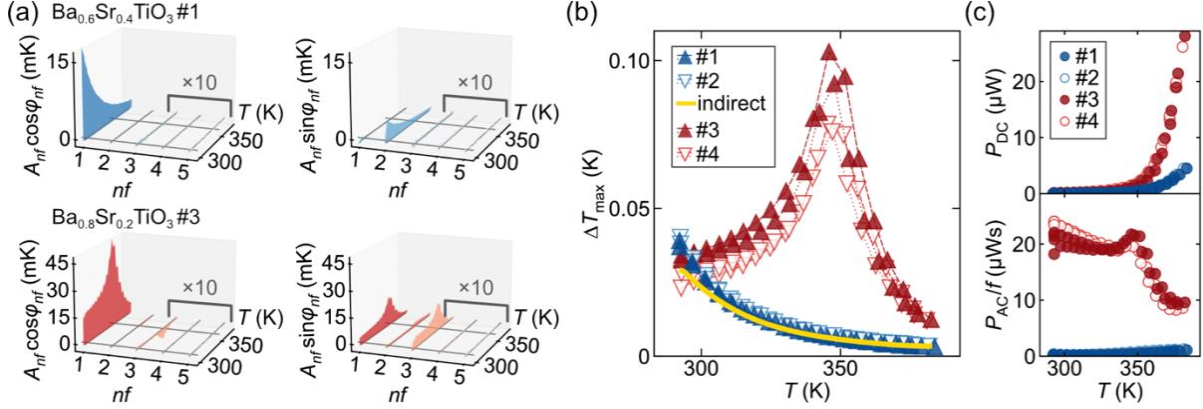

Fig. 3 (a) $nf$ and $T$ dependences of $A_{nf}$ and $\varphi_{nf}$, where in-phase and quadrature components are shown for completeness. (b) $T$ dependence of the temperature change $\Delta T_{max}$ generated when the electric field is varied from 0 to $E_{max}$=1.0 kV/mm, which corresponds to the ECE-induced adiabatic temperature change. The yellow solid line represents an estimation based on the indirect method with assuming $E$-independent specific heat. (c) DC ($P_{DC}$) and AC ($P_{AC}$) power consumptions measured by the integrated Sawyer-Tower circuit as a function of $T$.

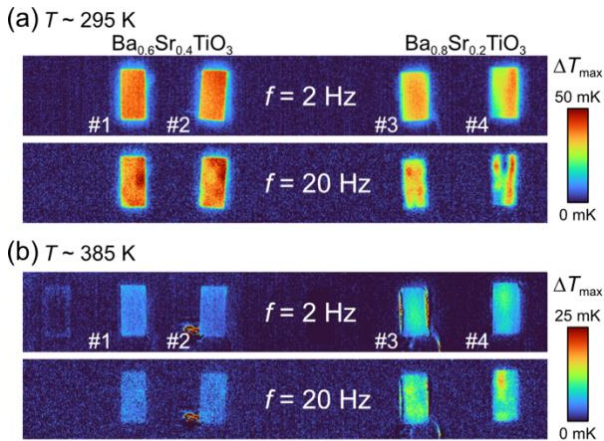

Fig. 4 $\Delta T_{max}$ images at $f$=2 and 20 Hz at $T$~295 K (a) and 385 K (b).

10various temperatures. (c) $E_{max}$ dependence of $A_{1f}$ and $\varphi_{1f}$ at various temperatures. The $T$, $A_{1f}$, and $\varphi_{1f}$ values were obtained by averaging the 25×55 values inside the samples [see the rectangular box in Fig. 1(c)].

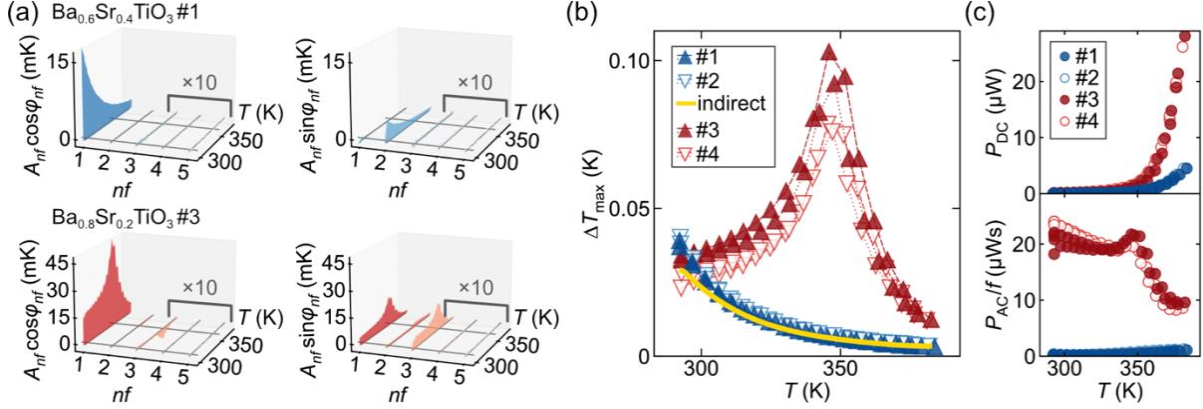

Fig. 3 (a) $nf$ and $T$ dependences of $A_{nf}$ and $\varphi_{nf}$, where in-phase and quadrature components are shown for completeness. (b) $T$ dependence of the temperature change $\Delta T_{max}$ generated when the electric field is varied from 0 to $E_{max}$=1.0 kV/mm, which corresponds to the ECE-induced adiabatic temperature change. The yellow solid line represents an estimation based on the indirect method with assuming $E$-independent specific heat. (c) DC ($P_{DC}$) and AC ($P_{AC}$) power consumptions measured by the integrated Sawyer-Tower circuit as a function of $T$.

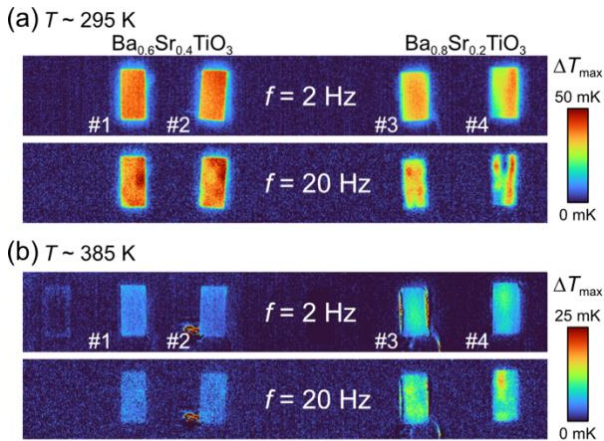

Fig. 4 $\Delta T_{max}$ images at $f$=2 and 20 Hz at $T$~295 K (a) and 385 K (b).

**Acknowledgements**

The authors thank M. Isomura for technical supports. This work was supported by JSPS KAKENHI Grant No. 20H02609, JST CREST "Creation of Innovative Core Technologies for Nano-enabled Thermal Management" Grant No. JPMJCR17I1, the Canon Foundation, and NIMS Joint Research Hub Program.
**References**

[1] A.S. Mischenko, Q. Zhang, J.F. Scott, R.W. Whatmore, and N.D. Mathur, Science **311**, 1270 (2006).
[2] B. Neese, B. Chu, S.-G. Lu, Y. Wang, E. Furman, and Q.M. Zhang, Science **321**, 821 (2008).
[3] J. Shi, D. Han, Z. Li, L. Yang, S.-G. Lu, Z. Zhong, J. Chen, Q.M. Zhang, and X. Qian, Joule **3**, 1200 (2019).
[4] X. Moya and N.D. Mathur, Science **370**, 797 (2020).
[5] R. Ma, Z. Zhang, K. Tong, D. Huber, R. Kornbluh, Y.S. Ju, and Q. Pei, Science **357**, 1130 (2017).
[6] E. Defay, R. Faye, G. Despesse, H. Strozyk, D. Sette, S. Crossley, X. Moya, and N.D. Mathur, Nat. Commun. **9**, 1827 (2018).
[7] L. Junjie, L. Jianting, Q. Shiqiang, S. Xiaopo, Q. Lijie, W. Yu, L. Turab, and B. Yang, Phys. Rev. Appl. **10**, 1 (2019).
[8] Y. Nouchokgwe, P. Lheritier, C.-H. Hong, A. Torelló, R. Faye, W. Jo, C.R.H. Bahl, and E. Defay, Nat. Commun. **12**, 3298 (2021).
[9] B. Nair, T. Usui, S. Crossley, S. Kurdi, G.G. Guzmán-Verri, X. Moya, S. Hirose, and N.D. Mathur, Nature **575**, 468 (2019).
[10] Y. Liu, J.F. Scott, and B. Dkhil, Appl. Phys. Rev. **3**, 031102 (2016).
[11] S. Crossley, B. Nair, R.W. Whatmore, X. Moya, and N.D. Mathur, Phys. Rev. X **9**, 041002 (2019).
[12] X. Moya, S. Kar-Narayan, and N.D. Mathur, Nat. Mater. **13**, 439 (2014).
[13] R. Niemann, O. Heczko, L. Schultz, and S. Fähler, Int. J. Refrig. **37**, 281 (2014).
[14] M. Marathe, A. Grünebohm, T. Nishimatsu, P. Entel, and C. Ederer, Phys. Rev. B **93**, 054110 (2016).
[15] Y. Liu, B. Dkhil, and E. Defay, ACS Energy Lett. **1**, 521 (2016).
[16] M. Baba, R. Kuwahara, N. Ishibashi, S. Fukuda, and M. Takeda, Rev. Sci. Instrum. **92**, 044902 (2021).
[17] O. Breitenstein, W. Warta, and M. Langenkamp, *Lock-in Thermography: Basics and Use for Evaluating Electronic Devices and Materials*, 2nd ed. (Springer-Verlag, Berlin Heidelberg, 2010).
[18] O. Wid, J. Bauer, A. Müller, O. Breitenstein, S.S.P. Parkin, and G. Schmidt, Sci. Rep. **6**, 28233 (2016).
[19] S. Daimon, R. Iguchi, T. Hioki, E. Saitoh, and K. Uchida, Nat. Commun. **7**, 13754 (2016).
[20] K. Uchida, S. Daimon, R. Iguchi, and E. Saitoh, Nature **558**, 95 (2018).
[21] Y. Hirayama, R. Iguchi, X.-F. Miao, K. Hono, and K. Uchida, Appl. Phys. Lett. **111**, 163901 (2017).
[22] R. Modak, R. Iguchi, H. Sepehri-Amin, A. Miura, and K. Uchida, AIP Adv. **10**, 065005 (2020).
[23] T. Hirai, R. Iguchi, A. Miura, and K. Uchida, Adv. Funct. Mater. **32**, 2201116 (2022).
[24] M.-H. Phan and S.-C. Yu, J. Magn. Magn. Mater. **308**, 325 (2007).
[25] V. Franco, J.S. Blázquez, J.J. Ipus, J.Y. Law, L.M. Moreno-Ramírez, and A. Conde, Prog. Mater. Sci. **93**, 112 (2018).
11

# Supplementary Material for
# Direct measurement of electrocaloric effect based on multi-harmonic lock-in thermography

Ryo Iguchi,[1,a)] Daisuke Fukuda,[2] Jun Kano,[2] Takashi Teranishi,[2,a)] and Ken-ichi Uchida,[1,3]

[1]*National Institute for Materials Science, Tsukuba 305-0047, Japan*

[2]*Graduate School of Natural Science and Technology, Okayama University, Okayama 700-8530, Japan*

[3]*Institute for Materials Research, Tohoku University, Sendai 980-8577, Japan*

[b)] Author to whom correspondence should be addressed: IGUCHI.Ryo@nims.go.jp and terani-t@cc.okayama-u.ac.jp

**Electrocaloric effect, heat losses, and Joule heating**

Here, we compare the temperature change due to the electrocaloric effect (ECE) and Joule heating based on a heat equation with heat loss to environment. With the heat loss term,[1] denoted by the time constant $\tau_s$, the heat equation is given by

$$C_v \frac{dT}{dt} = -T\left(\frac{\partial P}{\partial T}\right)\frac{dE}{dt} + q_J - \frac{(T-T_0)}{\tau_s}, \qquad (A2)$$

where $t$, $T$, $T_0$, $C_v$, $P$ and $E$ denote the time, system temperature, environment temperature, volumetric heat capacity, polarization, and electric field, respectively. $q_J = \sigma E^2$ denotes the Joule-heating induced power with $\sigma$ being the electrical conductivity. The complex temperature change $\delta T_{nf}$ at the harmonic frequency $nf$ can be calculated by substituting $E = E_{\max}(\mathrm{Im}[e^{i2\pi ft}] + 1)/2$ into Eq. (A1), where $f$ denotes the excitation frequency. Namely, it is determined by

$$\mathrm{Im}\left[\sum_{n=1}(i2\pi n f C_v + \tau_s^{-1})\delta T_{nf} e^{i2\pi nft}\right] = -\pi f T_0 E_{\max} \frac{\partial P(t)}{\partial T}\mathrm{Im}[i e^{i2\pi ft}] + q_J, \qquad (A3)$$

where $T = T_0 + \sum_n \mathrm{Im}[\delta T_{nf} e^{i2\pi nft}]$ and $\delta T_{nf} \ll T_0$. The heat loss effect due to $\tau_s$ affects both the ECE and Joule heating contributions but are clearly reduced if $f$ is much greater than the inverse of $\tau_s$.

Next, we discuss the behavior of ECE and the Joule heating. For simplicity, we focus on paraelectric systems where the linear relation $P = \varepsilon E$ holds and $\sigma$ is independent of $E$. In this condition, we obtain

$$\text{Im}\left[\sum_{n=1} (i2\pi n f C_v + \tau_s^{-1}) \delta T_{nf} e^{i2\pi n f t}\right]$$

$$= -\pi f \frac{T_0 E_{\max}^2}{2} \frac{\partial \varepsilon}{\partial T} \left(\text{Im}\left[\frac{e^{i4\pi f t}}{2} + i e^{i2\pi f t}\right]\right) + \frac{\sigma E_{\max}^2}{2} \left(\text{Im}\left[\frac{e^{i4\pi f t}}{4i} + e^{i2\pi f t}\right]\right), \quad (A4)$$

at 1$f$. This equation leads to

$$\delta T_{1f} = A_{1f} e^{-i\phi_{1f}} = \frac{1}{C_v + (i2\pi f \tau_s)^{-1}} \left(-T_0 \frac{\partial \varepsilon}{\partial T} + \frac{\sigma}{i\pi f}\right) \frac{E_{\max}^2}{4}, \quad (A5)$$

where $A_{1f}$ and $\varphi_{1f}$ denote the amplitude and phase delay at the elementary harmonic frequency, respectively. It can be found that, when the heat loss effect is negligibly small, the ECE term [the 1st term on the right-hand side of Eq. (A4)] induces the in-phase ($\varphi_{1f} = 0°$) and $f$ independent temperature oscillation, whereas the Joule heating term (the 2nd term) induces the out-of-phase one ($\varphi_{1f} = 90°$) with its magnitude decreasing with increasing $f$.